\documentclass[a4paper,11pt]{article}
\usepackage{pos}

\usepackage{graphicx}
\usepackage{verbatim}
\usepackage{slashed}
\usepackage{placeins}
\usepackage{setspace} 

\title{Mixing interactions and effects in the NJL-model}
\author*{Fabio L. Braghin}

\affiliation[a]{Instituto de F\'\i sica Universidade Federal de Goi\'as,\\
Av. Esperan\c ca, s/n,
 74690-900, Goi\^ania, GO, Brazil}

\emailAdd{braghin@ufg.br}

\abstract{
The flavor-dependent quark-antiquark  contact interactions,
induced by vacuum polarization and
recently derived for flavor U(3) Nambu-Jona-Lasinio model,
 are articulated with the resulting mixing effects
emerging from flavor symmetry breaking
in view. 
The formal effects of the explicit mixing interactions, $G_{i\neq j}$,  are detailed firstly for the 
meson mixing problem
without the inclusion of 't Hooft interactions induced by instantons.
Secondly, it is shown that these mixings, in the scalar channel of quark-antiquark interactions, 
might give rise to 
quark mixing in the gap equations.
Sixth order quark-antiquark interactions from vacuum polarization,
that break $U_A(1)$ symmetry,   also contribute.
}

\FullConference{The 11th International Workshop on Chiral Dynamics (CD2024)\\
 26-30 August 2024\\
Ruhr University Bochum, Germany
\\}

\begin{document}
\maketitle

\section{ Flavor dependent NJL coupling constants from polarization}

The constituent quark model provides 
a  successful description   of global hadron
properties. 
One of the most interesting and widespread relativistic versions
is found in the Nambu-Jona-Lasinio model \cite{NJL,NJL2,witten}.
This model incorporates naturally Dynamical Chiral Symmetry Breaking (DChSB)
with the corresponding multiplet of (almost) Goldstone bosons whose masses are not 
really zero due to the (small) symmetry breaking light quark masses  in
 the QCD Lagrangian that also provide the source
for flavor symmetry breaking (FSB) observed in hadron phenomenology
 \cite{FSB-QCD}. 
Light meson masses and weak decay constant are usually well described in this model in particular 
for pseudoscalar and vector mesons.
The chiral anomaly is usually incorporated by means of the 't Hooft determinantal interaction
induced by instantons  that lead to $2 N_f$ quark- effective interactions \cite{creutz}
and leads to the correct pseudoscalar 
meson (and vector meson)
nonet description with the $\pi^0-\eta-\eta'$ 
 mixing
\cite{meson-mixing,osipov}.
Flavor symmetry breaking  is also known to be  a source of meson mixings.
Overall, QCD low energy effective models must incorporate fundamental properties of QCD.
%The small  FSB  manifests in the non-degenerate light quark current masses 
%from QCD Lagrangian, and this is the only source of  FSB for effective models.
Accordingly, all the parameters of an effective model might
 depend on the FSB from QCD Lagrangian.
In the NJL model, the free parameters are the quark masses
 and the coupling constant, with 
an additional UV cutoff parameter. 
Whereas constituent quark masses are usually calculated by means of a one loop
level, via the
gap equation, the NJL coupling constant is simply considered as an external  parameter whose origin 
should be associated to gluon exchange in the long wavelength local limit.
%A contribution from the trace anomaly for  quark antiquark punctual  interactions has been
%found in \cite{EH-Seff}.
By considering  
the quark determinant with background quark currents, in one loop
approach,
it turns out that the resulting correction to the coupling constant, defined in the 
local long wavelength limit,  will depend on the FSB \cite{PRD-2021,JPG-2022}.
The resulting constituent quark - effective interaction
may be interpreted diagrammatically in terms of 
two gluon exchange (or one loop NJL four point Green function)
in spite of the need of redefining vertices.
This idea has been articulated in some works in the complete absence of the 't Hooft interaction
and of the explicit mixing interactions that emerge from FSB.
These flavor dependent interactions, even in the absence of 
explicit mixing interactions, lead to a quantum mixing in the sea quark dynamics - represented 
by the chiral condensates - that lead to tiny strange
contributions for the pion \cite{JPG-2022} (and for the kaon \cite{EPJA-2023}).
Structure of 
light scalar mesons present longstanding controverses that show some consonance
with predictions from the NJL-model \cite{JPG-2023}.
Concerning heavy mesons, NJL has shown to be not suitable along the years, although
 the use of non-covariant regularization with a three-dim momentum cutoff 
(whose value must be comparable  to,  or slightly larger than,  the QCD  cutoff, $\Lambda_{QCD}$) 
lead to surprisingly good description
of the 25-plet pseudoscalar
meson masses  and partially good (as it should be expected) 
of scalar mesons, although not appropriate for weak decay constants, \cite{EPJA-2023}.
In this work, few consequences of the explicit  mixing interactions obtained in this approach,
associated  to meson mixing interactions, are articulated and are shown to 
lead to quark mixing interactions.
In Refs. \cite{EPJA-2024,PRD-2014}, 
sixth order quark-antiquark $U_A(1)$ symmetry breaking 
 interactions have been derived from vacuum polarization
that must contribute for these mixings. 
Some of them  have the same shape of the 
't Hooft interactions for flavor U(3) \cite{creutz,osipov}
and they may also have some role in the mixing mechanisms.

The NJL-model with (normalized)  flavor dependent  coupling constants is given by
 \cite{PRD-2021,JPG-2022}:
\begin{eqnarray} \label{LeNJL}
{\cal L}_{eNJL} &=& \bar{\psi} S_{0,f}^{-1} \psi + 
(G_{ij} + \Delta G_{s,ij} )
( \bar{\psi} \lambda_i \psi ) ( \bar{\psi} \lambda_j \psi )
+  
G_{ij}  ( \bar{\psi} i\gamma_5 \lambda_i \psi ) ( \bar{\psi} i \gamma_5\lambda_j \psi ),
\end{eqnarray}
where $S_{0,f}^{-1} = i \slashed{\partial} - m_f$ for 
current quark masses $m_f$, $i,j = 0,1,... (N_f^2-1)$,
and the
 integral equations for the coupling constants have been presented in 
\cite{PRD-2021,JPG-2022} in terms of the quark propagators with
corrected mass  for the case of two gluon exchange obtained from the quark -determinant. 
In that case, these resulting coupling constants (and form factors) are UV finite.
For the one loop correction to the NJL model with NJL-interaction,
the corresponding flavor dependent interactions are UV divergent and require an 
UV cutoff.
A  renormalization procedure for 
the derivation of such effective interactions
was proposed in \cite{renormaliza}.
Coupling constants of the scalar and of the pseudoscalar channels
are different since one-loop process break chiral symmetry, although 
this difference has not yet been considered fully.
Some properties arise due to CP and U(1) symmetries:
$G_{11}=G_{22}$, $G_{44}=G_{55}$ and  $G_{66}=G_{77}$.
Mixing interactions arise solely for 
the neutral $i,j=0,3,8$ flavor states and depend on the quark constituent mass 
differences:
$G_{i \neq j} \propto (M_{f_1} - M_{f_2} )^n$, ${n=1,2}$.
%along the lines of 't Hooft interaction SU(3) at mean field.

Restricting  to the currents of the  diagonal flavor  generators, $i,j=0,3,8$,
the corresponding coupling constants 
can be written in terms of singlet quark currents,
 $G_{ff}$ ($f=u,d,s$) as:
\begin{eqnarray}  \label{gij-kf1f2}
G_{ij} (\bar{\psi} \lambda_i \psi ) 
 (\bar{\psi} \lambda_j \psi )  
= 2 \; G_{f_1f_2} (\bar{\psi}  \psi )_{f_1} 
 (\bar{\psi}   \psi )_{f_2} ,
\end{eqnarray}
By considering explicitely the mixing interactions  $G_{i\neq j}\neq 0$,
the  following relations between the coupling constants $G_{ii}$ and $G_{ff}$
 are obtained:
\begin{eqnarray} \label{G-K}
 2 G_{uu}
  &=&
 2  \frac{ G_{00}  }{3}
 + G_{33}  + \frac{G_{88} }{3} 
+ \frac{2\sqrt{2}}{3}  G_{08} + 2 \sqrt{\frac{2}{3}} G_{03} + \frac{2}{\sqrt{3}} G_{38}
= 2 I_{uu}
,
\nonumber
\\
2 G_{dd}
 &=& 2 \frac{ G_{00} }{3} 
 + G_{33}  + \frac{G_{88} }{3} 
+ \frac{2\sqrt{2}}{3}  G_{08} - 2 \sqrt{\frac{2}{3}} G_{03} - \frac{2}{\sqrt{3}} G_{38}
= 2 I_{dd}
,
\nonumber
\\
2 G_{ss} 
  &=& 2 \frac{ G_{00} }{3}
+ 4   \frac{G_{88} }{3}   - \frac{4\sqrt{2} }{3} G_{08}
= 2 I_{ss},
\end{eqnarray}
where  cancellations are seen to occur
and the momentum integrals $I_{uu} \propto \int \frac{d^4 k}{(2\pi)^4}
 S_{0,u}^* (k) S_{0,u} (k)$ are written in terms of each quark propagator.

Whereas interactions of the adjoint representation, $G_{ij}$, are 
responsible for the BSE amplitudes and corresponding meson structure,
 the 
interactions in the fundamental representation, $G_{f_1f_2}$, 
are responsible for the quark gap equations.
Note that  quark interactions $G_{ff}$ are independent of each other (flavor)
so that it may 
suggest there is 
no mixing in the gap equations without sixth order interactions.

\section{ BSE and mixings}

By means of the auxiliary field method \cite{AFM},
local meson fields are introduced for the scalar and pseudoscalar channels,
$S_i$ and $P_i$  (i=0,...8 for the flavor nonet).
%, such that
%the Lagrangian can be written as:
%\begin{eqnarray}
%{\cal L} = - \frac{i}{2} \sum_{ij} G_{ij}^{-1} P^i P^j 
%+ i Tr \ln ( S_{f,M}^{-1}  + \sum i \gamma_5 \lambda_i P_i )
%\end{eqnarray}
%where $S_{f,M}^{-1} = i \slashed{k} - M_f$ are  the quark propagators
%for constituent quark masses obtained from the coupled gap equations.
By considering the BSE, at the Born level, with flavor-dependent interactions 
the gap equations can be written as
\begin{eqnarray}
1 - 2 G_{ij} \Pi^{ij} = 0 ,
\end{eqnarray}
where there are mixings among states $i,j=0.3,8$ (i.e. $\pi^0-\eta-\eta'$ in the 
pseudoscalar channel or $\rho^0-\omega-\phi$ in the vector channel), so that
these BSE have the same shape of the NJL-model  equations with the use 
of the 't Hooft interactions at the mean field level \cite{NJL2}.
In this case, of complete account of the mixing interactions $G_{i\neq j}\neq 0$,
the gap equations remain uncoupled:
\begin{eqnarray}
M_f = m_f + i \; G_{ff} \; Tr S_{0,f}(0),
\end{eqnarray}
where $Tr$ is a generalized trace that includes integration in momentum
 of the quark propagator.

\section{ Gap equations and mixings}

The description of the pseudoscalar meson nonet 
may be understood by a rotation among the $P_3, P_0$ and $P_8$ 
auxiliary fields  defined for quark-antiquark meson states.
This rotation diagonalizes the  meson  nonet masses.
Similarly, by performing the same rotation in quark currents 
in the original Lagrangian \eqref{LeNJL},
for $i,j=0,3,8$, 
 the flavor dependent interactions of the adjoint representation
 become diagonal
and the  corresponding (explicit) mixing interactions disappear, i.e.
\begin{eqnarray}  \label{gij-kf1f2}
G_{ii} (\bar{\psi} \lambda_i \psi ) 
 (\bar{\psi} \lambda_i \psi )  
= 2 \; G_{f_1f_2} (\bar{\psi}  \psi )_{f_1} 
 (\bar{\psi}   \psi )_{f_2} .
\end{eqnarray}
This makes possible to 
solve BSE  for the whole nonet without mixings.
This is equivalent to, arbitrarily, switching off mixing interactions 
$G_{i\neq j}=0$, and it yields:
\begin{eqnarray} \label{G-K}
 2 G_{uu} = 2 G_{dd}
  &=&
  ( 14 I_{uu}  + 14 I_{dd} + 8 I_{ss})/18
,
\nonumber
\\
2 G_{ss} 
  &=& ( 8 I_{uu}  + 8  I_{dd} + 20 I_{ss})/18  ,
\nonumber
\\
2 G_{ud} &=&
( - 8 I_{uu}  - 8  I_{dd} + 16 I_{ss})/18 ,
\nonumber
\\
2 G_{us} = 2 G_{ds} &=&
( 4 I_{uu} + 4  I_{dd} - 8  I_{ss})/18 .
\end{eqnarray}
It is seen that there are strange (sea quarks) components in the 
u - and d - quark  self interactions.

Although the BSE are now diagonal, the quark interactions 
have mixings $G_{f_1 \neq f_2} \neq 0$, and these 
yield the following form for the gap equations:
\begin{eqnarray}
M_f^* = m_f + i \; G_{f f_2} \; Tr S_{0, f_2} (0),
\end{eqnarray}
that  has an implicit sum in $f_2$ and that
presents a very similar shape to the gap equations obtained from the 
't Hooft interactions at the mean field level.
The resulting mixing interactions in the fundamental representation
are proportional to the quark mass differences
$G_{f_1 f_2} \propto (M_{f_1} - M_{f_2})^n$ for $n=1,2$ in the leading order.
Therefore they are much smaller than the diagonal ones $G_{ff}$ for the light quarks. 

\section{ Mixing from sixth order quark-antiquark interactions}

Sixth order quark interactions produced by vacuum polarization 
have been derived in \cite{EPJA-2024,PRD-2014} for flavor U(3) and U(2)
that can be written in terms of quark currents for a particular Dirac and flavor channel.
In spite of the large variety of momentum dependent 
couplings that may arise, there are some non derivative couplings
that include at least one scalar current.
They can be written as \cite{EPJA-2024}:
\begin{eqnarray} 
\label{6th-S-PS}
{\cal L}_{6,1} &=& 
 G_{sb,ps} T^{ijk}  
( J_i^S J_j^{PS} J_k^{PS} + 
 J_i^{PS} J_j^{PS} J_k^S
+  J_i^{PS} J_j^S J_k^{PS} ) 
-
\; G_{sb,s} T^{ijk}  
J_i^S J_{j}^S J_k^S 
,
\\
\label{6th-V-A}
{\cal L}_{6,2} &=&
 T^{ijk}  G_{sb1}   \left[ 
J^{\mu}_{V,i} J_{\mu}^{V,j}    J_{S,k}
+
   \frac{G_{sb2}}{ G_{sb1} }  J^{\mu}_{A,i} J_{\mu}^{A,j}
  J_{S,k}  \right],
%- i
%\left(   
%J^{\mu}_{V,i} J_{\mu}^{A,j}  J_{PS,k}  
%- i
% J^{\mu}_{A,i} J_{\mu}^{V,j}
% J_{PS,k}
% \right)
% \right],
\end{eqnarray}
where $T^{ijk} = 2 (d_{ijk} + i f_{ijk})$ with both  SU(3) structure constants,
complemented with the $i,j=0$ components, 
and
the coupling constants were calculated as zero external momentum limit of  
 one-loop form factors in \cite{EPJA-2024}.
Note that  the first line \eqref{6th-S-PS} presents interactions with the same shape of the 
't Hooft interactions for flavor SU(3) although the scalar and pseudoscalar sectors have 
different coupling constants.
By resorting to  a mean field approximation,
 along the lines
usually  performed for the 	
't Hooft interactions,
 the scalar currents can provide
mixing interactions.
\begin{eqnarray}
J_0^S    &\to&  
< \bar{\psi} \lambda_0 \psi > =
\sqrt{\frac{2}{3}}
\left( < \bar{u} u >   + < \bar{d} d >    +  < \bar{s} s > \right), 
\nonumber
\\
J_3^S    &\to&   < \bar{\psi} \lambda_3 \psi > = 
  < \bar{u} u >  -  < \bar{d} d >     , 
\nonumber
\\
J_8^S    &\to&   < \bar{\psi} \lambda_8 \psi >  = 
\frac{1}{\sqrt{3}}
\left( < \bar{u} u >   + < \bar{d} d >    -  2  < \bar{s} s > \right).
\end{eqnarray}
These reductions make possible to write fourth order 
quark-antiquark interactions by considering the 6th order interactions above, on the example of the 
't Hooft interactions.
These interactions can be written as:
\begin{eqnarray}
\label{Gij-6th}
 {\cal L}_{mix}^{6}
&=&
G_{ij}^{6,ps}  J_i^{PS}  J^{PS}_j
+
G_{ij}^{6,s}  J^S_i J^S_j
+
G_{ij}^{6,v}
J_{V,i}^{\mu}  J^{V,j}_{\mu}
+ 
G_{ij}^{6,a}
J^{A,i}_{\mu}  J_{A,j}^{\mu},
\end{eqnarray}
where
\begin{eqnarray}
G_{ij}^{6,ps}  &=& 
6 d_{jik}   {G}_{sb,ps}  
< \bar{\psi} \lambda_k \psi >,
\;\;\;\;\;
G_{ij}^{6,s}  = 
 - 6
 d_{jik}  {G}_{sb,s}  
< \bar{\psi} \lambda_k \psi >,
\nonumber
\\
G_{ij}^{6,v}   &=& 
6  d_{jik}
 {G}_{sb1} 
< \bar{\psi} \lambda_k \psi > ,
\;\;\;\;\;
G_{ij}^{6,a}   =
6   d_{jik}
 {G}_{sb2} .
< \bar{\psi} \lambda_k \psi > 
.
\end{eqnarray}
Some preliminary numerical estimations have been provided in \cite{EPJA-2024} to which contributions from
't Hooft interactions  can be added.

\section{ Summary }

Fourth order flavor-dependent mixing interactions for the NJL-model
were shown to provide explicit contributions for 
meson mixings and for quark mixings when analyzed, respectively, in 
the adjoint representation and in the fundamental representation when a diagonalization of 
neutral pseudoscalar
(meson) states is done.
Meson  mixings, as usual, appear in the coupled BSE 
that, by diagonalization, lead to mixing quark-antiquark interactions.
Interactions with the same shape of the 't Hooft interactions also emerge from 
vacuum polarization, although
 there is an intrinsic ambiguity in comparing their 
relative strengths.
Relevance of this mechanism for the spectrum of charmed pseudoscalar mesons 
in a covariant picture
\cite{SBK}
and to  aspects of pseudoscalar meson decay 
are  intended to be exploited soon.
Numerical estimations, complementing those presented in 
\cite{PRD-2021,JPG-2022,JPG-2023,EPJA-2023,EPJA-2024}, by taking into account 't Hooft interactions,
%and more extensive articulation of the ambiguity(ies) in comparing
%the strength of  these interactions with 
%the 't Hooft interactions
will be provided elsewhere.

\section*{Acknowledgements}

 The author thanks short conversations  with C. Weiss and L. Gan.
F.L.B. is a member of
INCT-FNA,  Proc. 464898/2014-5.
F.L.B. acknowledges partial support from 
 CNPq-312750/2021 and CNPq-407162/2023-2.

\end{document}